\documentclass[english,preprint,superscriptaddress]{revtex4}
\usepackage[T1]{fontenc}
\usepackage{amsmath}
\usepackage{amssymb}
\usepackage{graphicx}
\usepackage{esint}

\usepackage{amsmath}
\usepackage{amssymb}
\usepackage{graphicx}
\usepackage{esint}

\makeatletter

\providecommand{\tabularnewline}{\\}

\usepackage{amsthm}\usepackage{latexsym}\usepackage{bm}\usepackage{amsfonts}

\usepackage{hyperref}

\@ifundefined{showcaptionsetup}{}{%
 \PassOptionsToPackage{caption=false}{subfig}}
\usepackage{subfig}
\makeatother

\begin{document}

\title{Explicit High-Order Gauge-Independent Symplectic Algorithms for Relativistic
Charged Particle Dynamics}

\author{Jianyuan Xiao}
\affiliation{Department of Engineering and Applied Physics, University of Science and Technology of China, Hefei, 230026, China}

\author{Hong Qin}
\affiliation{Department of Engineering and Applied Physics, University of Science and Technology of China, Hefei, 230026, China}
\affiliation{Plasma Physics Laboratory, Princeton University, Princeton, NJ 08543, U.S.A}
\email{hongqin@ustc.edu.cn}

\begin{abstract}
Symplectic schemes are powerful methods for numerically integrating
Hamiltonian systems, and their long-term accuracy and fidelity have
been proved both theoretically and numerically. However direct applications
of standard symplectic schemes to relativistic charged particle dynamics
result in implicit and electromagnetic gauge-dependent algorithms.
In the present study, we develop explicit high-order gauge-independent
noncanonical symplectic algorithms for relativistic charged particle
dynamics using a Hamiltonian splitting method in the 8D phase space.
It also shown that the developed algorithms can be derived as variational
integrators by appropriately discretizing the action of the dynamics.
Numerical examples are presented to verify the excellent long-term
behavior of the algorithms.
\end{abstract}



\maketitle
\global\long\def\EXP{\times10^}  \global\long\def\rmm{\mathrm{m}}  \global\long\def\rmc{\mathrm{c}}  \global\long\def\rms{\mathrm{s}}  \global\long\def\rmd{\mathrm{d}}  \global\long\def\diag{\textrm{diag}}  \global\long\def\xs{ \mathbf{x}_{s}}  \global\long\def\bfx{\mathbf{x}}  \global\long\def\bfv{\mathbf{v}}  \global\long\def\bfp{\mathbf{p}}  \global\long\def\bfA{\mathbf{A}}  \global\long\def\bfB{\mathbf{B}}  \global\long\def\bfS{\mathbf{S}}  \global\long\def\bfG{\mathbf{G}}  \global\long\def\bfE{\mathbf{E}}  \global\long\def\bfM{\mathbf{M}}  \global\long\def\bfQ{\mathbf{Q}}  \global\long\def\bfu{\mathbf{u}}  \global\long\def\bfe{\mathbf{e}}  \global\long\def\bfd{\mathbf{d}}  \global\long\def\rme{\mathrm{e}}  \global\long\def\rmi{\mathrm{i}}  \global\long\def\rmq{\mathrm{q}}  \global\long\def\ope{\omega_{pe}}  \global\long\def\oce{\omega_{ce}}  \global\long\def\FIG#1{Fig.~\ref{#1}}  \global\long\def\EQ#1{Eq.~(\ref{#1})}  \global\long\def\SEC#1{Sec.~\ref{#1}}  \global\long\def\APP#1{Appendix~\ref{#1}}  \global\long\def\REF#1{Ref.~\cite{#1}}  \global\long\def\DFDDELTATAU#1{\frac{#1_{l+1}-#1_l}{\Delta\tau}}  \global\long\def\DFDDELTATAUP#1{\frac{#1_{l}-#1_{l-1}}{\Delta\tau}}  \global\long\def\DDELTAT#1{\textrm{Dt}\left(#1\right)}  \global\long\def\DDELTATA#1{\textrm{Dt}^*\left(#1\right)}  \global\long\def\GRADD{ {\mathrm{\nabla_{d}}}}  \global\long\def\CURLD{ {\mathrm{curl_{d}}}}  \global\long\def\DIVD{ {\mathrm{div_{d}}}}  \global\long\def\CURLDP{ {\mathrm{curl_{d}}^{T}}}  \global\long\def\cpt{\captionsetup{justification=raggedright }}  \global\long\def\act{\mathcal{A}}  \global\long\def\calL{\mathcal{L}}  \global\long\def\calJ{\mathcal{J}}  \global\long\def\DELTAA{\left( \bfA_{J,l}-\bfA_{J,l}' \right)}  \global\long\def\DELTAAL{\left( \bfA_{J,l-1}-\bfA_{J,l-1}' \right)}  \global\long\def\ADAGGER{\bfA_{J,l}^\dagger}  \global\long\def\ADAGGERA#1{\bfA_{J,#1}^{x/2}}  \global\long\def\EDAGGER#1{\bfE_{J,#1}^{x/2}}  \global\long\def\BDAGGER#1{\bfB_{J,#1}^{x/2}}  \global\long\def\DDT{\frac{\partial}{\partial t}}  \global\long\def\DBYDT{\frac{\rmd}{\rmd t}}  \global\long\def\DBYANY#1{\frac{\partial }{\partial #1}} \newcommand{\WZERO}[1]{W_{\sigma_0 I}\left( #1 \right)} \newcommand{\WONE}[1]{W_{\sigma_1 J}\left( #1 \right)} \newcommand{\WONEJp}[1]{W_{\sigma_1 J'}\left( #1 \right)} \newcommand{\WTWO}[1]{W_{\sigma_2 K}\left( #1 \right)} \global\long\def\bfzig{\mathbf{r}_{\textrm{zig2}}}   \global\long\def\bfxzig{\mathbf{r}_{\textrm{xzig}}}   \global\long\def\bfzzig{\mathbf{r}_{\textrm{zzig}}}   \global\long\def\xzig{\mathbf{x}_{\textrm{zig}}}   \global\long\def\yzig{\mathbf{y}_{\textrm{zig}}}   \global\long\def\zzig{\mathbf{z}_{\textrm{zig}}}   \global\long\def\zigspmvar{\left( \bfx_{sp,l-1},\bfx_{sp,l},\tau \right)}   \global\long\def\zigspvar{\left( \bfx_{sp,l},\bfx_{sp,l+1},\tau \right)}   \global\long\def\MQQ{M_{00}} \global\long\def\MDQDQ{M_{11}} \global\long\def\MDQQ{M_{01}}  

\section{Introduction}

Charged particle dynamics plays an important role in plasma physics,
space physics and accelerator physics. In a given electromagnetic
field, the dynamics of a charged particle is described by Newton's
equation with the Lorentz force. Since the governing equation is a
6D nonlinear ordinary differential equation (ODE) in general, we have
to depend on numerical solutions to understand the complicated behavior
of the dynamics. In practice, long-term simulations are often needed.
For instance in a typical tokamak, the particle confinement time of
ions is $10^{7}\sim10^{8}$ times longer than their cyclotron period.
For these multi-scale dynamics, it is crucial to adopt numerical schemes
with the long-term conservation properties. Conventional integrators
for ODEs, such as the 4th order Runge-Kutta (RK4) method, can bound
the truncation error of the discrete time advance for each time step.
However these truncation errors from different time-steps will accumulate
during the simulation and the global error grows without bound.

Fortunately, most physical systems are Hamiltonian, and symplectic
(or geometric) integrators for Hamiltonian systems have been systematically
studied since 1980s \cite{Lee82,Ruth83,Feng85,Feng86,Lee87,Veselov88,yoshida1990construction,Forest90,Channell90,Candy91,Tang93,Sanz-Serna94,Shang99,marsden2001discrete,Hairer02,Feng10}.
The idea of symplectic integrators is to construct time advance maps
that preserve the symplectic 2-form, just as the exact solutions of
the original Hamiltonian system do. It has been demonstrated that
symplectic integrators can globally bound the errors on the invariants
of the dynamics \cite{Ruth83,Feng86,Hairer02}, such as the conserved
Hamiltonian and momenta, for all simulation time-steps. 

Recently in plasma physics and accelerator physics, various symplectic
algorithms have been developed and applied for the Vlasov-Maxwell
system, Vlasov-Poisson system \cite{squire2012geometric,xiao2013variational,kraus2013variational,evstatiev2013variational,Shadwick14,xiao2015variational,xiao2015explicit,Qin15JCP,he2015hamiltonian,qin2016canonical,Webb16,kraus2017gempic,xiao2018structure},
two-fluid dynamics \cite{xiao2016explicit}, magnetohydrodynamics
\cite{gawlik2011geometric,pavlov2011structure,zhou2014variational,zhou2015formation},
and guiding center dynamics \cite{qin2008variational,qin2009variational,guan2010phase,li2011variational,zhang2014,squire2012gauge,burby2017toroidal,ellison2018degenerate}.
For charged particle dynamics in a given electromagnetic field, the
dependence of the Hamiltonian $H(p,q)$ on momentum $p$ and position
$q$ is inseparable in general, and direct applications of standard
symplectic methods will result in implicit schemes. Recently, He et
al. \cite{he2015explicit,he2017explicit} discovered a Hamiltonian
splitting method to build explicit high-order symplectic algorithms
for non-relativistic charged particle dynamics in static electromagnetic
fields, and its applicability has been extended to general electromagnetic
fields and relativistic dynamics in the canonical setting \cite{zhou2017explicit}.
Generating function methods have also been utilized to construct explicit
3rd order symplectic method for relativistic dynamics \cite{zhang2016explicit,zhang2018explicit}.
As an example of another class of geometric integrators, the well
known Boris algorithm \cite{boris1972proceedings} was found to preserve
phase volume \cite{qin2013boris}, but not the symplectic structure
\cite{ellison2015comment}. Families of volume preserving algorithms
have been developed for relativistic and non-relativistic charged
particle dynamics \cite{zhang2015volume,he2015volume,he2016high,He16-172,Tu2016,Higuera2017}. 

In the present study, we develop a family of explicit high-order gauge-independent
noncanonical symplectic integrators for relativistic charged particle
dynamics using the Hamiltonian splitting method discovered by He et
al. \cite{he2015explicit,he2017explicit}. The algorithms possess
desirable properties for long-term simulation studies of relativistic
charged particle dynamics. For example, it preserves a noncanonical
symplectic 2-form that enables the global bound on errors for invariants
of the dynamics. Because the algorithms are explicit, higher accuracy
can be achieved with relatively low computational cost. The gauge-independent
property implies that discrete orbits are not affected by the choice
of electromagnetic gauge. Compared with the algorithms in Ref. \cite{zhou2017explicit},
the methods developed in the present study do not require the knowledge
of vector and scalar potentials. Only electromagnetic fields are needed.
We will also show that the noncanonical symplectic algorithms developed
can be derived as variational integrators with specifically constructed
discrete Lagrangian.

The paper is organized as follows. In \SEC{SecTHEORY}, we start
from the Lagrange theory of the relativistic charged particle dynamics,
and derive the corresponding noncanonical Hamiltonian theory and Poisson
bracket. In \SEC{SecALGO}, explicit high-order gauge-independent
noncanonical symplectic integrators are constructed using the Hamiltonian
splitting method. The same schemes are also derived as variational
integrators. Numerical examples are given in \SEC{SecNE}.

\section{Lagrangian and noncanonical Hamiltonian formalism of relativistic
charged particle dynamics}

\label{SecTHEORY} The motion of a relativistic charged particle in
a given electromagnetic fields is governed by Newton's equation with
the Lorentz force,
\begin{eqnarray}
\frac{\rmd\bfx}{\rmd t} & = & \frac{\bfp}{\gamma}~,\label{1}\\
\frac{\rmd\bfp}{\rmd t} & = & \bfE+\frac{\bfp\times\bfB}{\gamma}~,\label{2}
\end{eqnarray}
where $\gamma=\sqrt{1+\left|\bfp\right|^{2}}$ is the relativistic
factor. For simplicity, the rest mass $m_{0}$, speed of light $\rmc$
and charge of the particle $q$ are set to be 1. The Lagrangian theory
for relativistic particle dynamics can be found in Ref. \cite{goldstein2011classical}.
In the present study, we adopt the Lagrangian theory in the 8D tangent
bundle of space-time. The proper time $\tau$ is used as the parameter
for particle's worldline in the 8D tangent bundle. The Lagrangian
$L$ and action integral $S$ are 
\begin{eqnarray}
L[\bfx,t] & = & \frac{1}{2}\left(-\dot{t}^{2}+|\dot{\bfx}|^{2}\right)+\dot{\bfx}\cdot\bfA\left(\bfx,t\right)-\dot{t}\phi\left(\bfx,t\right)~,\\
S[\bfx,t] & = & \int L[\bfx,t]\rmd\tau~,
\end{eqnarray}
where $\bfA$ and $\phi$ are vector and scalar potentials. Particle's
space-time coordinates $\bfx$ and $t$ are functions of the proper
time $\tau$, and $\dot{\mathbf{x}}=d\mathbf{x}/d\tau$ and $\dot{t}=dt/d\tau$.
The governing equations are the Euler-Lagrange equations, 
\begin{eqnarray}
\frac{\delta S}{\delta\bfx} & = & 0~,\label{5}\\
\frac{\delta S}{\delta t} & = & 0~.\label{6}
\end{eqnarray}
If we let $\bfp=\dot{\bfx}$, it can be proved that Eqs.\,\eqref{5}
and \eqref{6} are exactly the same as Eqs.\,\eqref{1} and \eqref{2}.

To obtain the noncanonical Hamiltonian theory, we need to derive the
Lagrange 1-form $\Gamma$ \cite{marsden2013introduction} defined
as 
\begin{eqnarray}
\Gamma=\frac{\partial L}{\partial\dot{\bfx}}\bfd\bfx+\frac{\partial L}{\partial\dot{t}}\bfd t~.
\end{eqnarray}
where $\bfd$ denotes for the exterior derivative. The Euler-Lagrange
equation can be written as 
\begin{eqnarray}
i_{(\dot{\bfx},\dot{\bfp},\dot{t},\dot{\gamma})}\bfd\Gamma+\bfd H=0~,\label{EqnITDGAMMA}
\end{eqnarray}
where $\bfp=\dot{\bfx}$ and $(\dot{\bfx},\dot{\bfp},\dot{t},\dot{\gamma})$
denotes for the following vector field in the 8D cotangent bundle,
\begin{eqnarray}
\dot{\bfx}\frac{\partial}{\partial\bfx}+\dot{\bfp}\frac{\partial}{\partial\bfp}+\dot{t}\frac{\partial}{\partial t}+\dot{\gamma}\frac{\partial}{\partial\gamma}~
\end{eqnarray}
In Eq.\,\eqref{EqnITDGAMMA}, $H$ is the Hamiltonian 
\begin{eqnarray}
H & = & \frac{\partial L}{\partial\dot{\bfx}}\cdot\dot{\bfx}-L~\nonumber \\
 & = & \frac{1}{2}\left(-\gamma^{2}+\left|\bfp\right|^{2}\right)~.\label{EqnHAMT}
\end{eqnarray}
Since $H$ does not depend on $\tau,$ $H$ is an invariant of the
dynamics, which implies that the particle is always on the mass-shell.
Equation \eqref{EqnITDGAMMA} can be also written in matrix form, 

\begin{equation}
\dot{z}\Omega=-\frac{\partial}{\partial z}H\left(z\right)~,
\end{equation}
or

\begin{eqnarray}
\dot{z} & = & \Omega^{-1}\left(\frac{\partial}{\partial z}H\left(z\right)\right)^{T}~,\label{EqnHamtZ}
\end{eqnarray}
where $\Omega$ is the matrix form of the noncanonical symplectic
2-form $\bfd\Gamma$, and $z=\left(\bfx,\bfp,t,\gamma\right)$ is
a point in the 8D cotangent bundle. It is clear that \EQ{EqnHamtZ}
is a noncanonical Hamilton's equation
\begin{eqnarray}
\dot{z} & = & \left\{ z,H\right\} ~,\label{EqnHamPoitZ}
\end{eqnarray}
with a noncanonical Poisson bracket $\left\{ .,.\right\} $. Specifically,
the noncanonical Poisson bracket is defined by $\Omega^{-1}$ as 
\begin{eqnarray}
\left\{ F,G\right\}  & = & \left(\frac{\partial}{\partial z}F\right)\Omega^{-1}\left(\frac{\partial}{\partial z}G\right)^{T}\\
 & = & \left(\frac{\partial}{\partial z}F\right)\left[\begin{array}{cccc}
0 & I & 0 & 0\\
-I & \hat{B}\left(\bfx,t\right) & 0 & -\bfE\left(\bfx,t\right)^{T}\\
0 & 0 & 0 & -I\\
0 & \bfE\left(\bfx,t\right) & I & 0
\end{array}\right]\left(\frac{\partial}{\partial z}G\right)^{T}~,\\
\hat{B} & = & \left[\begin{array}{ccc}
0 & B_{z} & -B_{y}\\
-B_{z} & 0 & B_{x}\\
B_{y} & -B_{x} & 0
\end{array}\right]\,.
\end{eqnarray}
It can be verified that \EQ{EqnHamPoitZ} is equivalent to the following
dynamic equation 
\begin{eqnarray}
\left\{ \begin{array}{ccl}
\dot{\bfx} & = & \bfp~,\\
\dot{\bfp} & = & \gamma\bfE+\bfp\times\bfB~,\\
\dot{t} & = & \gamma~,\\
\dot{\gamma} & = & \bfE\cdot\bfp~.
\end{array}\right.\label{EqnDDp}
\end{eqnarray}
which reduces to Eqs.\,\eqref{1} and \eqref{2}.

\section{Construction of the geometric algorithm}

\label{SecALGO} In previous works, the powerful Hamiltonian splitting
technique has been applied to render explicit high-order symplectic
algorithms for single particle dynamics \cite{he2015explicit,he2017explicit,zhou2017explicit},
Vlasov-Maxwell systems \cite{xiao2015explicit,he2015explicit,he2016hamiltonian,kraus2017gempic},
and two-fluid dynamics \cite{xiao2016explicit}. Here, we apply a
similar technique to the noncanonical Hamilton's equation \eqref{EqnHamPoitZ}
for relativistic particle dynamics. The Hamiltonian $H$ in \EQ{EqnHAMT}
can be naturally split into four parts, 
\begin{eqnarray}
H & = & H_{t}+H_{x}+H_{y}+H_{z}~,\\
H_{t} & = & -\gamma^{2}/2\thinspace,\\
H_{x} & = & p_{x}^{2}/2\,,\\
H_{y} & = & p_{y}^{2}/2\,,\\
H_{z} & = & p_{z}^{2}/2\thinspace.
\end{eqnarray}
For $H_{t}$, Hamilton's equation is
\begin{eqnarray}
\dot{z} & = & \left\{ z,H_{t}\right\} ~,
\end{eqnarray}
i.e.,
\begin{eqnarray}
\left\{ \begin{array}{ccl}
\dot{\bfx} & = & 0~,\\
\dot{\bfp} & = & \gamma\bfE~,\\
\dot{t} & = & \gamma~,\\
\dot{\gamma} & = & 0~.
\end{array}\right.
\end{eqnarray}
Its exact solution map $\Theta_{t}\left(\Delta\tau\right)$ is 
\begin{eqnarray}
\Theta_{t}\left(\Delta\tau\right):\left\{ \begin{array}{ccl}
\bfx & \rightarrow & \bfx~,\\
\bfp & \rightarrow & \bfp+\Delta\tau\gamma\int_{0}^{1}\rmd t'\bfE\left(\bfx,t+\gamma\Delta\tau t'\right)~,\\
t & \rightarrow & t+\Delta\tau\gamma~,\\
\gamma & \rightarrow & \gamma~.
\end{array}\right.
\end{eqnarray}
Exact solution maps for the subsystems $H_{x}$, $H_{y}$ and $H_{z}$
can be obtained similarly. For $H_{i}$, $i\in\left\{ x,y,z\right\} $,
Hamilton's equation is 
\begin{eqnarray}
\left\{ \begin{array}{ccl}
\dot{\bfx} & = & p_{i}\bfe_{i}~,\\
\dot{\bfp} & = & p_{i}\bfe_{i}\times\bfB~,\\
\dot{t} & = & 0~,\\
\dot{\gamma} & = & E_{i}p_{i}~,
\end{array}\right.
\end{eqnarray}
and the solution map is 
\begin{eqnarray}
\Theta_{i}\left(\Delta\tau\right):\left\{ \begin{array}{ccl}
\bfx & \rightarrow & x+\Delta\tau p_{i}\bfe_{i}~,\\
\bfp & \rightarrow & \bfp+\Delta\tau\int_{0}^{1}\rmd t'p_{i}\bfe_{i}\times\bfB\left(\bfx+p_{i}\bfe_{i}\Delta\tau t',t\right)~,\\
t & \rightarrow & t~,\\
\gamma & \rightarrow & \gamma+\Delta\tau p_{i}\int_{0}^{1}E_{i}\left(\bfx+p_{i}\bfe_{i}\Delta\tau t',t\right)\rmd t'~.
\end{array}\right.
\end{eqnarray}
Using these exact solutions of subsystems, we can construct high-order
explicit algorithms by various compositions. Since exact solutions
are symplectic, the algorithms constructed by composition are automatically
symplectic. For example, a 1st order symplectic scheme is
\begin{eqnarray}
\Theta_{1}\left(\Delta\tau\right)=\Theta_{x}\left(\Delta\tau\right)\Theta_{y}\left(\Delta\tau\right)\Theta_{z}\left(\Delta\tau\right)\Theta_{t}\left(\Delta\tau\right)~,\label{EqnFSTHS}
\end{eqnarray}
and a symmetric 2nd order symplectic scheme can be built using Strang
splitting \cite{Hairer02},
\begin{eqnarray}
\Theta_{2}\left(\Delta\tau\right) & = & \Theta_{x}\left(\Delta\tau/2\right)\Theta_{y}\left(\Delta\tau/2\right)\Theta_{z}\left(\Delta\tau/2\right)\Theta_{t}\left(\Delta\tau\right)\nonumber \\
 &  & \Theta_{z}\left(\Delta t/2\right)\Theta_{y}\left(\Delta t/2\right)\Theta_{x}\left(\Delta t/2\right)~.\label{EqnHAMS2}
\end{eqnarray}
A $2(l+1)$-th order scheme can be constructed from a $2l$-th order
scheme using the method of triple jump \cite{yoshida1990construction,Hairer02},
\begin{eqnarray}
\Theta_{2(l+1)}(\Delta\tau) & = & \Theta_{2l}(\alpha_{l}\Delta\tau)\Theta_{2l}(\beta_{l}\Delta\tau)\Theta_{2l}(\alpha_{l}\Delta\tau)~,\\
\alpha_{l} & = & 1/(2-2^{1/(2l+1)})~,\\
\beta_{l} & = & 1-2\alpha_{l}~.
\end{eqnarray}
The main difficulty in implementing the present algorithm is calculating
integrals in each solution map. When these integrals can not be calculated
explicitly, we can approximate the external electromagnetic fields
$\bfB$ and $\bfE$ by piece-wise polynomial fields $\bar{\bfB}$
and $\bar{\bfE}$ that satisfy Maxwell's equation. For example in
vacuum, they satisfy \cite{xiao2015explicit} 
\begin{eqnarray}
\dot{\bar{\bfB}} & = & -\nabla\times\bar{\bfE}~,\\
0 & = & \nabla\cdot\bar{\bfB}~.
\end{eqnarray}
The piece-wise polynomial approximation can be made to arbitrary high-orders.

We have found previously that the explicit high-order noncanonical
symplectic particle-in-cell (PIC) scheme can be also obtained by using
the discrete variational method \cite{xiao2018structure}. The same
idea applies here, i.e., the present noncanonical relativistic particle
integrators can be derived as variational integrators \cite{marsden1998multisymplectic,marsden2001discrete,Hairer02}.
For this purpose, we consider a 1st order approximation of discrete
action integral 
\begin{eqnarray}
S_{d1} & = & \sum_{l=0}^{N_{t}-1}L_{d1}\left(x_{l}^{4},x_{l+1}^{4};\Delta\tau\right)~,
\end{eqnarray}
where $L_{d1}\left(x_{l}^{4},x_{l+1}^{4};\Delta\tau\right)$ is the
discrete Lagrangian 
\begin{eqnarray}
L_{d1}\left(x_{l}^{4},x_{l+1}^{4};\Delta\tau\right) & = & \frac{1}{2}\left(-\left(\frac{t_{l+1}-t_{l}}{\Delta\tau}\right)^{2}+\left|\frac{\bfx_{l+1}-\bfx_{l}}{\Delta\tau}\right|^{2}\right)+\nonumber \\
 &  & \DFDDELTATAU x\int_{0}^{1}\rmd\tau'A_{x}\left(x_{l}+\left(x_{l+1}-x_{l}\right)\tau',y_{l},z_{l},t_{l}\right)+\nonumber \\
 &  & \DFDDELTATAU y\int_{0}^{1}\rmd\tau'A_{y}\left(x_{l+1},y_{l}+\left(y_{l+1}-y_{l}\right)\tau',z_{l},t_{l}\right)+\nonumber \\
 &  & \DFDDELTATAU z\int_{0}^{1}\rmd\tau'A_{z}\left(x_{l+1},y_{l+1},z_{l}+\left(z_{l+1}-z_{l}\right)\tau',t_{l}\right)-\nonumber \\
 &  & \DFDDELTATAU t\int_{0}^{1}\rmd\tau'A_{t}\left(x_{l+1},y_{l+1},z_{l+1},t_{l}+\left(t_{l+1}-t_{l}\right)\tau'\right).
\end{eqnarray}
Here, $x_{l}^{4}$ represents $\left(x_{l},y_{l},z_{l},t_{l}\right)$
and $A_{t}=\phi$ is the scalar potential. Discrete equation of motion
can be derived by the discrete variational principle,
\begin{eqnarray}
\frac{\partial S_{d1}}{\partial\bfx_{l}} & = & 0~,\label{EqnDSODL}\\
\frac{\partial S_{d1}}{\partial t_{l}} & = & 0~,\label{EqnDSODLT}
\end{eqnarray}
for $1\leq l\leq N_{t}$. Written out explicitly, Eq.\,\eqref{EqnDSODLT}
is
\begin{eqnarray}
 &  & -\frac{t_{l+1}-2t_{l}+t_{l-1}}{\Delta\tau^{2}}+\DFDDELTATAU x\int_{0}^{1}\rmd\tau'A_{x,t}\left(x_{l}+\left(x_{l+1}-x_{l}\right)\tau',y_{l},z_{l},t_{l}\right)+\nonumber \\
 &  & \DFDDELTATAU y\int_{0}^{1}\rmd\tau'A_{y,t}\left(x_{l+1},y_{l}+\left(y_{l+1}-y_{l}\right)\tau',z_{l},t_{l}\right)+\nonumber \\
 &  & \DFDDELTATAU z\int_{0}^{1}\rmd\tau'A_{z,t}\left(x_{l+1},y_{l+1},z_{l}+\left(z_{l+1}-z_{l}\right)\tau',t_{l}\right)-\nonumber \\
 &  & \DFDDELTATAU t\int_{0}^{1}\rmd\tau'\left(1-\tau'\right)A_{t,t}\left(x_{l+1},y_{l+1},z_{l+1},t_{l}+\left(t_{l+1}-t_{l}\right)\tau'\right)+\nonumber \\
 &  & \frac{1}{\Delta\tau}\int_{0}^{1}\rmd\tau'A_{t}\left(x_{l+1},y_{l+1},z_{l+1},t_{l}+\left(t_{l+1}-t_{l}\right)\tau'\right)-\nonumber \\
 &  & \DFDDELTATAUP t\int_{0}^{1}\rmd\tau'\tau'A_{t,t}\left(x_{l},y_{l},z_{l},t_{l-1}+\left(t_{l}-t_{l-1}\right)\tau'\right)-\nonumber \\
 &  & \frac{1}{\Delta\tau}\int_{0}^{1}\rmd\tau'A_{t}\left(x_{l},y_{l},z_{l},t_{l-1}+\left(t_{l}-t_{l-1}\right)\tau'\right)~.\label{EqnDSDT}
\end{eqnarray}
Let $\bfE=-\nabla A_{t}-\partial\bfA/\partial t=\left(E_{1}\left(x^{4}\right),E_{2}\left(x^{4}\right),E_{3}\left(x^{4}\right)\right)$,
$\bfp_{l}=\DFDDELTATAU\bfx=[p_{1,l},p_{2,l},p_{3,l}]$, and $\gamma=\DFDDELTATAU t$.
Using the following identities, 
\begin{eqnarray*}
\frac{\rmd}{\rmd\tau'}f\left(\bfx+p_{i}\bfe_{i}\tau'\Delta\tau,t\right) & = & p_{i}\Delta\tau\tau'f_{,i}\left(\bfx+p_{i}\bfe_{i}\tau'\Delta\tau,t\right)~,\\
\frac{\rmd}{\rmd\tau'}\left(\tau'f\left(\bfx,t+\gamma\tau'\Delta\tau\right)\right) & = & \gamma\tau'\Delta\tau f_{,t}\left(\bfx,t+\gamma\tau'\Delta\tau\right)+f\left(\bfx,t+\gamma\tau'\Delta\tau\right)~,\\
\frac{\rmd}{\rmd\tau'}\left(\left(1-\tau'\right)f\left(\bfx,t+\gamma\tau'\Delta\tau\right)\right) & = & \gamma\Delta\tau\left(1-\tau'\right)f_{,t}\left(\bfx,t+\gamma\tau'\Delta\tau\right)-f\left(\bfx,t+\gamma\tau'\Delta\tau\right)~,
\end{eqnarray*}
we can rewrite \EQ{EqnDSDT} as 
\begin{eqnarray}
\DFDDELTATAUP\gamma & = & \sum_{i=1}^{3}p_{i,l}\int_{0}^{1}\rmd t'E_{i}\left(\bfx_{l}+\sum_{j<i\textrm{ and }1\leq j\leq3}p_{j,l}\bfe_{j}\Delta\tau+p_{i,l}\Delta\tau t',t_{l}\right)~,\label{EqnDSFDT}
\end{eqnarray}
which is an explicit scheme for advancing $\gamma_{l}$. A similar
treatment applies to Eq.\,\eqref{EqnDSODL} as well, leading to
\begin{eqnarray}
\DFDDELTATAUP\bfp & = & \int_{0}^{1}\rmd\tau'\gamma_{l}\bfE\left(\bfx_{l},t_{l}+\gamma_{l}\tau'\right)+\bfp_{l-1}\cdot\hat{\bfB}_{p,l-1}+\bfp_{l}\cdot\hat{\bfB}_{p,l}^{*}~,\label{EqnDSFDP}
\end{eqnarray}
where 
\begin{equation}
\bfp_{l-1}\cdot\hat{\bfB}_{p,l-1}=\left[\begin{array}{c}
p_{y,l-1}\int_{0}^{1}\rmd t'B_{z,l-1}\left(x_{l},y_{l-1}+t'p_{y,l-1}\Delta\tau,z_{l-1},t_{l-1}\right)-\\
p_{z,l-1}\int_{0}^{1}\rmd t'B_{y,l-1}\left(x_{l},y_{l},z_{l-1}+t'p_{z,l-1}\Delta\tau,t_{l-1}\right)~,\\
p_{z,l-1}\int_{0}^{1}\rmd t'B_{x,l-1}\left(x_{l},y_{l},z_{l-1}+t'p_{z,l-1}\Delta\tau,t_{l-1}\right)~,\\
0
\end{array}\right]~,
\end{equation}
\begin{equation}
\bfp_{l}\cdot\hat{\bfB}_{p,l}^{*}=\left[\begin{array}{c}
0,\\
-p_{x,l}\int_{0}^{1}\rmd t'B_{z,l}\left(x_{l}+t'p_{x,l}\Delta\tau,y_{l},z_{l},t_{l}\right)~,\\
p_{x,l}\int_{0}^{1}\rmd t'B_{y,l}\left(x_{l}+t'p_{x,l}\Delta\tau,y_{l},z_{l},t_{l}\right)-\\
p_{y,l}\int_{0}^{1}\rmd t'B_{x,l}\left(x_{l+1},y_{l}+t'p_{y,l}\Delta\tau,z_{l},t_{l}\right)~
\end{array}\right]~.
\end{equation}
Equation (\ref{EqnDSFDP}) furnishes an explicit scheme for advancing
$\bfx_{l}$. It can be seen that Eqs. (\ref{EqnDSFDT}) and (\ref{EqnDSFDP})
are the same as $\Theta_{1}$ in \EQ{EqnFSTHS}.

For higher order splitting schemes, equivalent variational integrators
also exist. For example, the discrete action integral from which a
scheme equivalent to $\Theta_{2}$ can be derived is 
\begin{eqnarray}
S_{d2} & = & \sum_{l=0}^{N_{t}-1}\Delta tL_{d2}\left(x_{2l}^{4},x_{2l+1}^{4},x_{2l+2}^{4};\Delta\tau\right)~,\label{EqnActS2}
\end{eqnarray}
where 
\begin{eqnarray*}
L_{d2}\left(x_{2l}^{4},x_{2l+1}^{4},x_{2l+2}^{4};\Delta\tau\right) & = & L_{d1}\left(x_{2l}^{4},x_{2l+1}^{4};\Delta\tau/2\right)+L_{d1}'\left(x_{2l+1}^{4},x_{2l+2}^{4};\Delta\tau/2\right)~,\\
L_{d1}'\left(x_{l}^{4},x_{l+1}^{4};\Delta\tau\right) & = & \frac{1}{2}\left(-\left(\frac{t_{l+1}-t_{l}}{\Delta\tau}\right)^{2}+\left|\frac{\bfx_{l+1}-\bfx_{l}}{\Delta\tau}\right|^{2}\right)+\\
 &  & -\DFDDELTATAU t\int_{0}^{1}\rmd\tau'A_{t}\left(x_{l},y_{l},z_{l},t_{l}+\left(t_{l+1}-t_{l}\right)\tau'\right)+\\
 &  & \DFDDELTATAU z\int_{0}^{1}\rmd\tau'A_{z}\left(x_{l},y_{l},z_{l}+\left(z_{l+1}-z_{l}\right)\tau',t_{l+1}\right)+\\
 &  & \DFDDELTATAU y\int_{0}^{1}\rmd\tau'A_{y}\left(x_{l},y_{l}+\left(y_{l+1}-y_{l}\right)\tau',z_{l+1},t_{l+1}\right)+\\
 &  & \DFDDELTATAU x\int_{0}^{1}\rmd\tau'A_{x}\left(x_{l}+\left(x_{l+1}-x_{l}\right)\tau',y_{l+1},z_{l+1},t_{l+1}\right)~.
\end{eqnarray*}

The gauge-independent property can be directly shown from the form
of the discrete Lagrangian. If we change potentials $\bfA$ and $\phi$
by a gauge field $\psi$ in the discrete action $S_{d1}$ as 
\begin{eqnarray}
\bfA & \rightarrow & \bfA+\nabla\psi~,\\
\phi & \rightarrow & \phi-\frac{\partial\psi}{\partial t}~,
\end{eqnarray}
 $S_{d1}$ is changed only by a boundary term 
\begin{eqnarray}
S_{d1}\rightarrow S_{d1}-\psi\left(x_{0}^{4}\right)+\psi\left(x_{N_{t}}^{4}\right)~.
\end{eqnarray}
Thus the evolution determined by Eqs. (\ref{EqnDSODL}) and (\ref{EqnDSODLT})
is independent of the gauge field $\psi$.

\section{Numerical Examples}

\label{SecNE} We have implemented the explicit 2nd order Gauge-Independent
Geometric Integrator (GIGI2) for relativistic particle dynamics. In
this section, we test the performance of the GIGI2 using several numerical
examples, can compare it with the RK4 method.

\subsection{The 2D Tokamak Geometry\label{subsec:The-2D-Tokamak}}

\begin{figure}
\subfloat[GIGI2]{\includegraphics[width=0.49\textwidth]{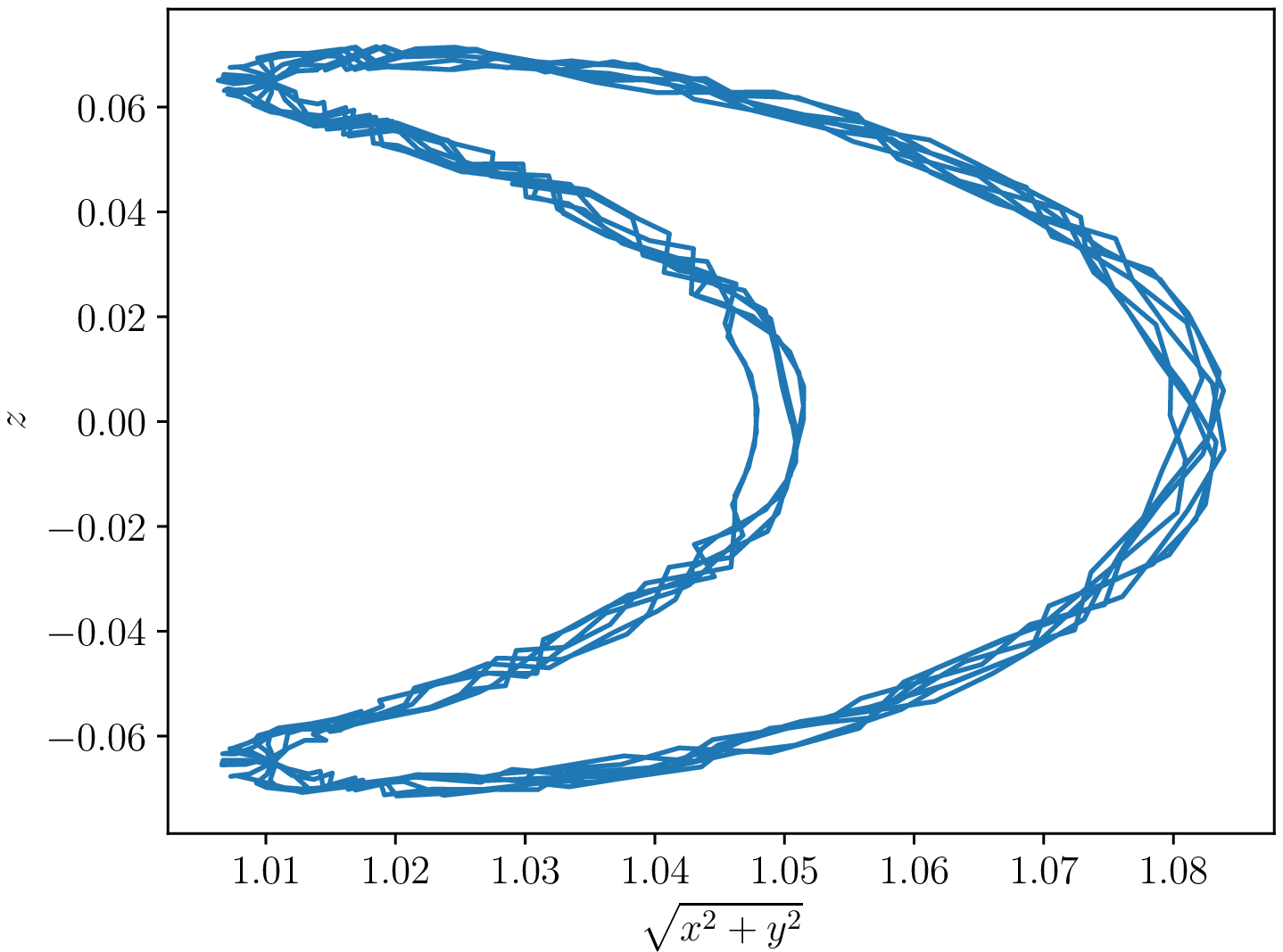}

}\subfloat[RK4]{\includegraphics[width=0.49\textwidth]{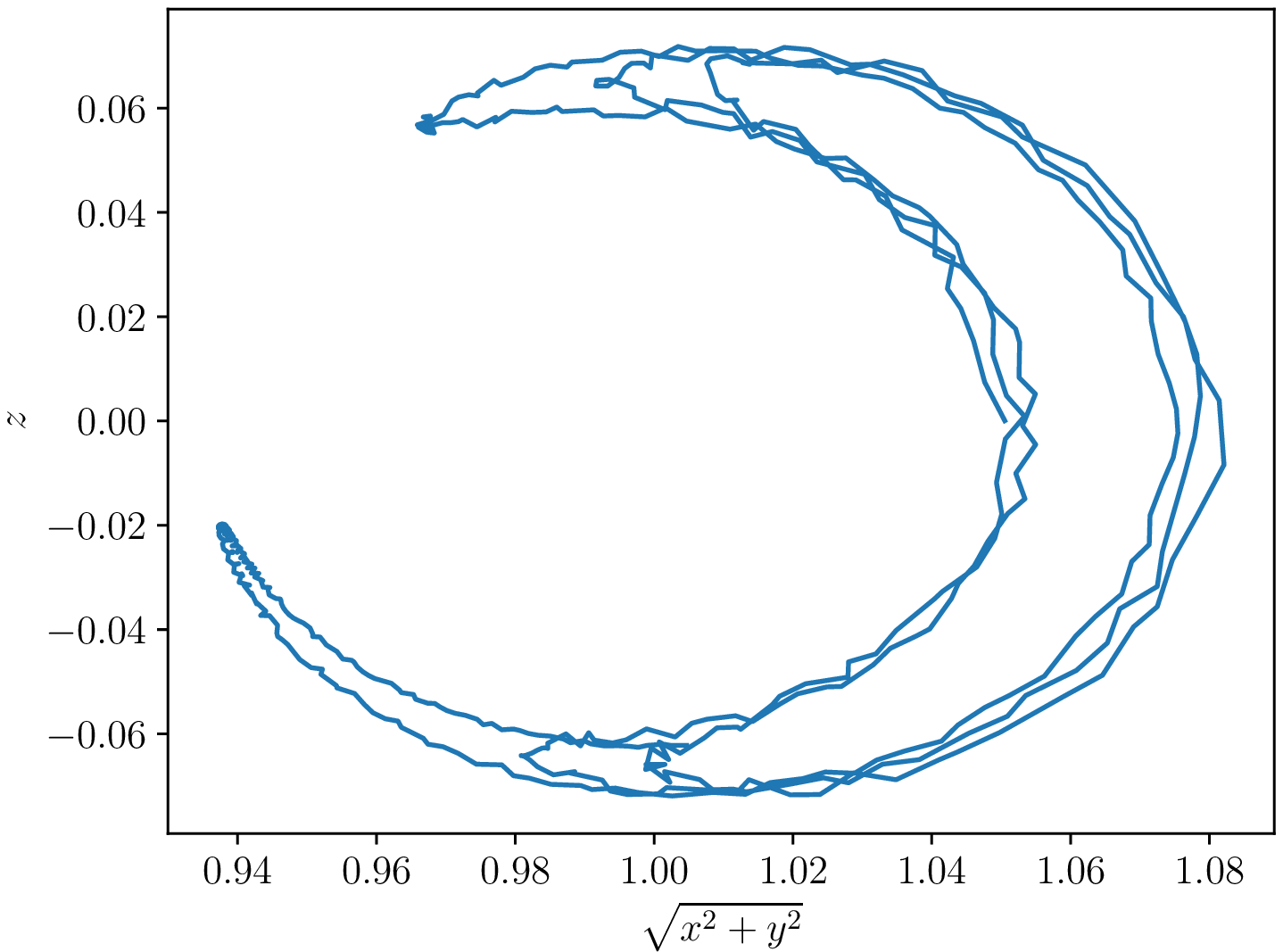}

}

\caption{Particle orbit in the poloidal plane of a tokamak obtained by the
GIGI2 and RK4 method.}
\label{FigTOKAMAK} 
\end{figure}

\begin{figure}
\begin{centering}
\includegraphics[width=0.6\textwidth]{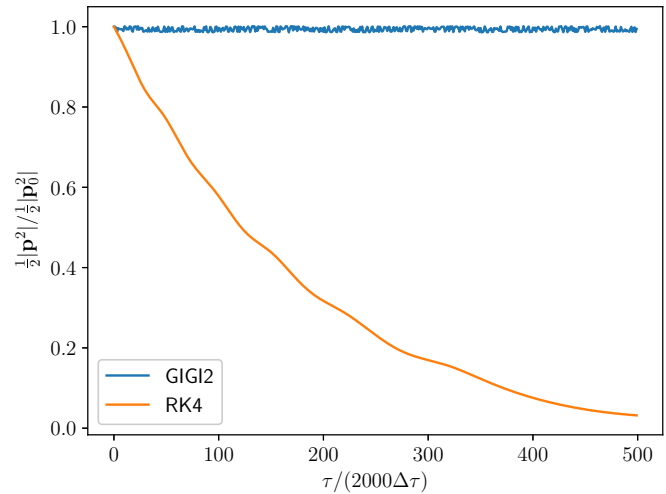} 
\par\end{centering}
\caption{Evolution of $\frac{1}{2}|\bfp|^{2}$ of the charged particle in a
tokamak calculated by the GIGI2 and RK4 method.}
\label{FigTOKAMAKE} 
\end{figure}

The first example is the dynamics of a charged particle in a 2D tokamak
geometry. The magnetic potential and electrostatic potential are 
\begin{eqnarray}
\bfA\left(x,y,z,t\right) & = & B_{0}\left(\frac{r^{2}}{2R}\bfe_{\xi}-\frac{\log(R/R_{0})R_{0}}{2}\bfe_{z}+\frac{R_{0}z}{2R}\bfe_{R}\right)~,\\
\phi\left(x,y,z,t\right) & = & 0~,
\end{eqnarray}
where 
\begin{eqnarray}
R & = & \sqrt{x^{2}+y^{2}}~,\\
r & = & \sqrt{\left(R-R_{0}\right)^{2}+z^{2}}~,\\
\bfe_{\xi} & = & [-\frac{y}{R},\frac{x}{R},0]~,\\
\bfe_{R} & = & [\frac{x}{R},\frac{y}{R},0]~,
\end{eqnarray}
and $B_{0}$ is the strength of the magnetic field at $R=R_{0}\text{ and }z=0$.
The normalization of physical quantities in numerical calculation
is listed in Tab.~\ref{TabNORM}.
\begin{table}
\begin{centering}
\begin{tabular}{|c|c|c|}
\hline 
Names & Symbols & Units\tabularnewline
\hline 
\hline 
Position & $\bfx,r,R$ & $\rmc/\Omega$\tabularnewline
\hline 
Time & $t,\tau$ & $1/\Omega$\tabularnewline
\hline 
Momentum & $\bfp$ & $m_{0}\rmc$\tabularnewline
\hline 
Velocity & $\bfv$ & $\rmc$\tabularnewline
\hline 
Magnetic field & $\bfB$ & $m_{0}\Omega/q$\tabularnewline
\hline 
Electric field & $\bfE$ & $\rmc m_{0}\Omega/q$\tabularnewline
\hline 
\end{tabular}
\par\end{centering}
\caption{Normalization used in the numerical example of Sec.~\ref{SecNE}.
Here, $\Omega=qB_{0}/m_{0}$.}
\label{TabNORM}
\end{table}
After the normalization, the magnetic field is 

\begin{equation}
\bfB=\frac{r}{2R}\bfe_{\theta}+\frac{R_{0}\Omega}{\rmc R}\bfe_{\xi}~,
\end{equation}
and the motion equation of the particle is exactly Eq. (\ref{EqnDDp}).
We set $R_{0}\Omega/\rmc=1$, and initially the particle is located
at $\bfx_{0}=[1.05,0,0]$ and its velocity is $\bfv_{0}=[2.1\EXP{-3},4.3\EXP{-4},0]$.
The time step is set to be $\Delta\tau=0.25$, and the total number
of time steps is $1\EXP{6}$. During the simulation the location and
energy are recorded, and results are plotted in Figs.~\ref{FigTOKAMAK}
and \ref{FigTOKAMAKE}. It is evident that the GIGI2 method preserves
the orbit and energy well, whereas the RK4 method does not. 

\subsection{Accelerator Field\label{subsec:Accelerator-Field}}

The second example is a charged particle in a model linear accelerator
configuration with 
\begin{eqnarray}
\bfA\left(x,y,z,t\right) & = & \frac{B_{0}}{2a}\sin\left(k_{z}z\right)\left(x^{2}-y^{2}\right)\bfe_{z}~,\\
\phi\left(x,y,z,t\right) & = & \phi_{0}\sin\left(k_{r}z-\omega t\right)~.
\end{eqnarray}
Here, $\bfA$ is the periodic quadrupole focusing field in the transverse
direction, $\phi$ provides the accelerating radio frequency (RF)
field in the longitudinal direction, and $a$ is the radius of the
transverse direction. The normalization of physical variables used
in the calculation is the same as that listed in Tab.~\ref{TabNORM}.
The normalized external electromagnetic fields are 

\begin{eqnarray}
	\bfB\left(x,y,z,t\right) & = & \frac{\rmc}{a\Omega}\sin\left(k_{z}\rmc z/\Omega\right)\left(y\bfe_{x}-x\bfe_{y}\right)~,\label{EqnB}\\
\bfE\left(x,y,z,t\right) & = & -\frac{\Omega\phi_{0}q}{m_{0}\rmc}k_{r}\cos\left(k_{r}\rmc z/\Omega-\omega t/\Omega\right)\bfe_{z}~.\label{EqnE}
\end{eqnarray}

\begin{figure}
\begin{centering}
\subfloat[GIGI2]{\includegraphics[width=0.49\textwidth]{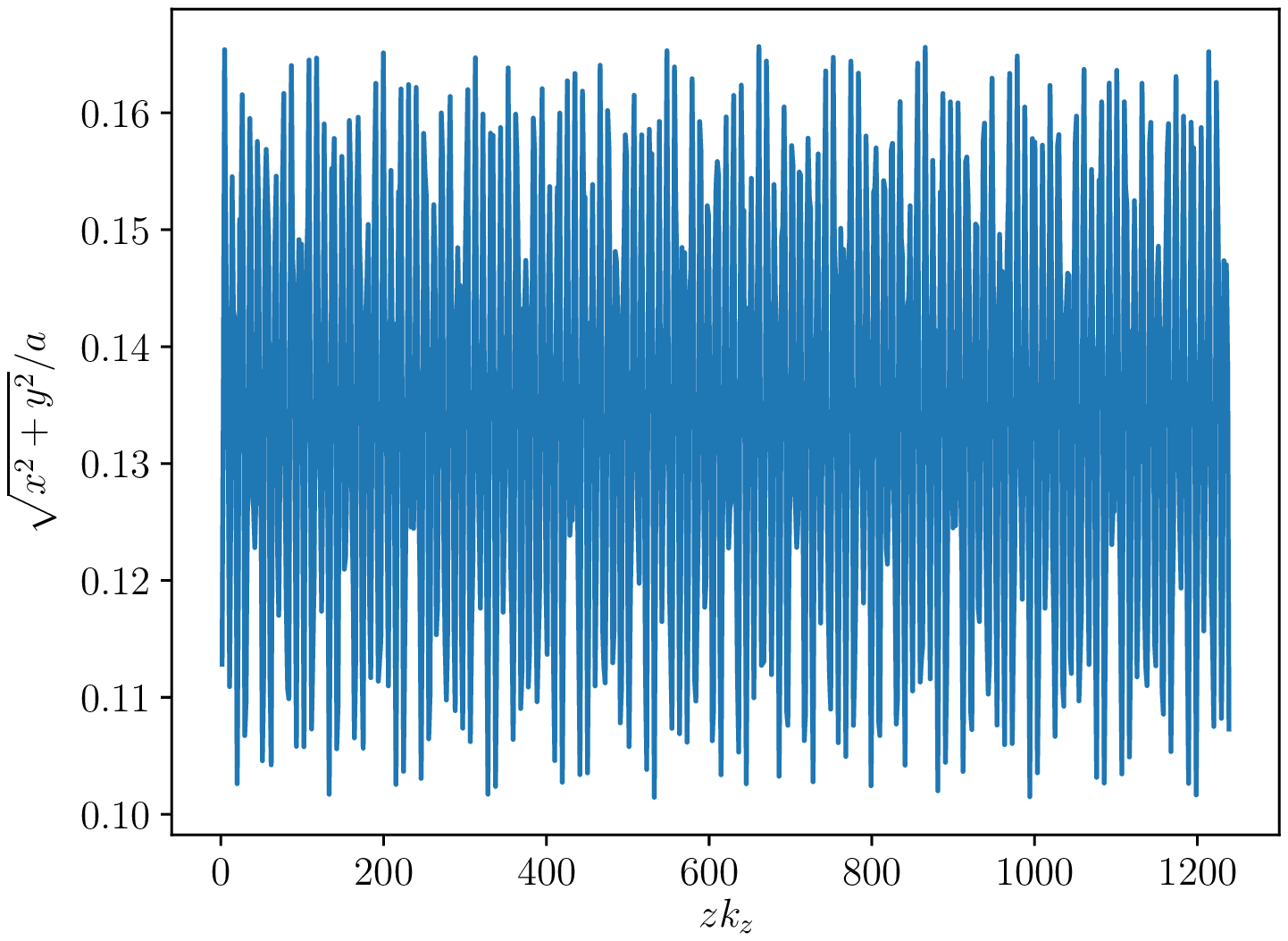}

}\subfloat[RK4]{\includegraphics[width=0.49\textwidth]{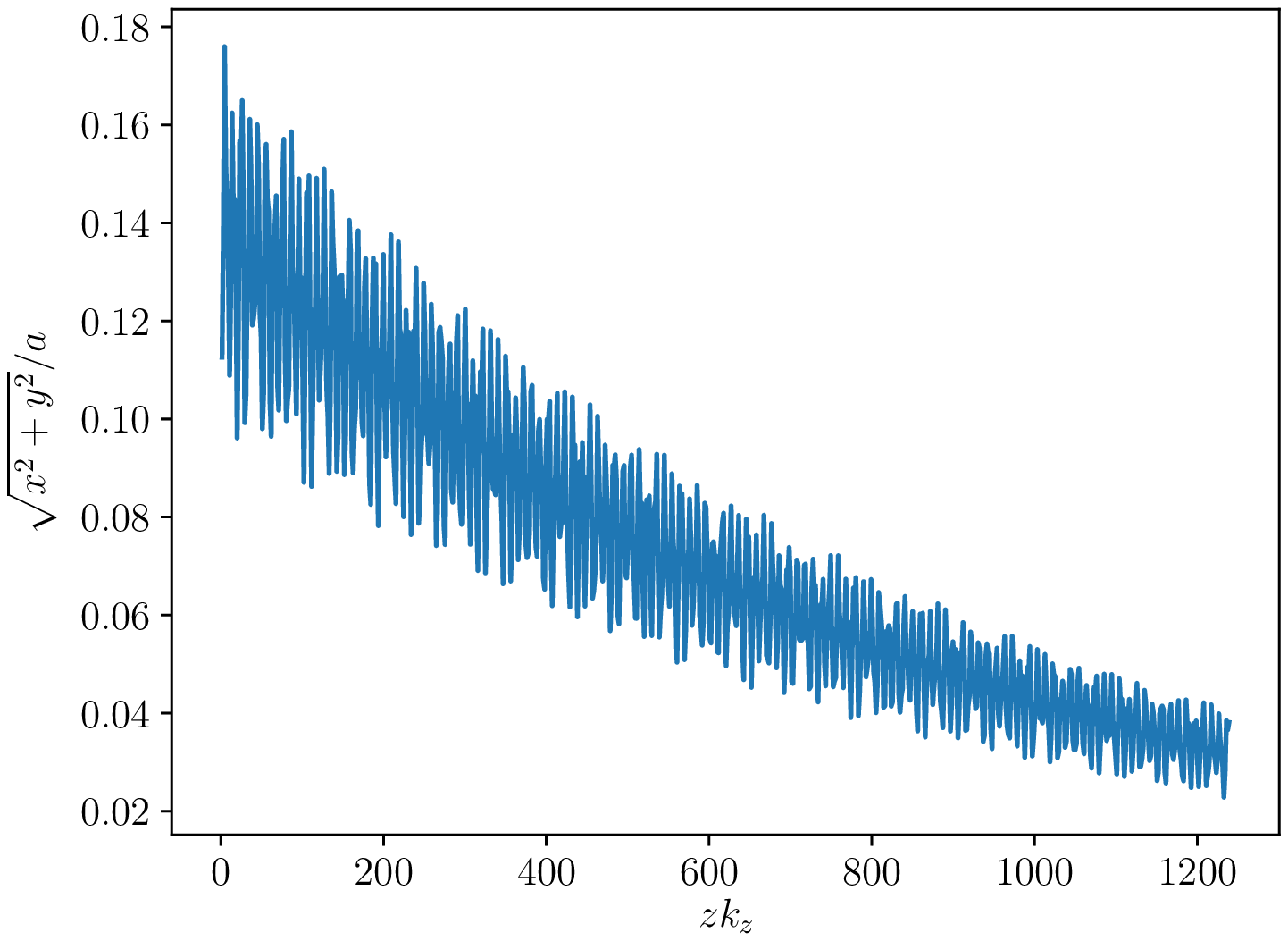}

}
\par\end{centering}
\caption{Particle orbit in a model accelerator simulated by the GIGI2 (a) and
RK4 (b) method.}
\label{FigACCORB} 
\end{figure}

\begin{figure}
\begin{centering}
\includegraphics[width=0.6\textwidth]{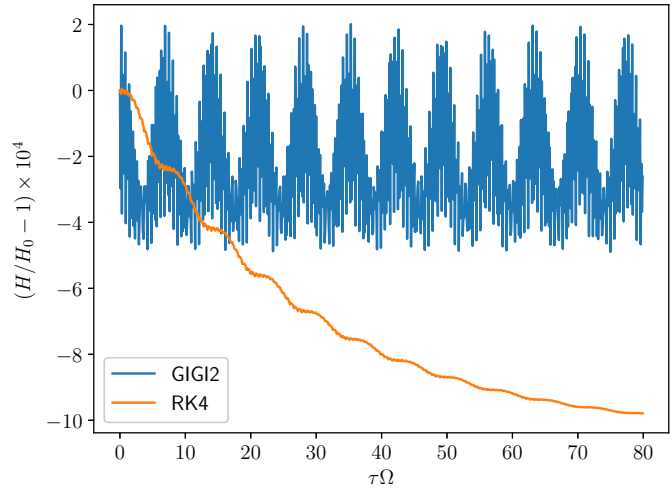} 
\par\end{centering}
\caption{Evolution of the Hamiltonian $H$ of the particle in a model accelerator
calculated by the GIGI2 and RK4 method.}
\label{FigACCENE} 
\end{figure}

First, the longitudinal accelerating field $\bfE$ is turned off,
and particle's dynamics in the quadrupole focusing lattice is examined.
Simulation parameters are chosen as 
\begin{eqnarray*}
\frac{\rmc}{a\Omega} & = & 30~,\\
k_{z}\rmc z/\Omega&=& 7.5,\\
\bfx_{t=0} & = & [2.667\EXP{-3},2.667\EXP{-3},0]~,\\
\left.\frac{\rmd\bfx}{\rmd t}\right|_{t=0} & = & [-0.001,0.001,0.9]~,\\
\Delta\tau & = & 0.1~.
\end{eqnarray*}
The total number of time steps is 800. Particle's orbit and error
on the Hamiltonian are plotted in Figs. \ref{FigACCORB} and \ref{FigACCENE}.
It is observed that particle's orbit obtained by the GIGI2 is stable,
and particle dynamics in the transverse direction is the betatron
oscillation, as expected \cite{Davidson01-all,Qin13PRL2,Qin14-044001,Qin15-056702}
. The error on the Hamiltonian is globally bounded by a small number
for the GIGI2. On the other hand, the RK4 method fails to generate
the correct orbit, and its error on the Hamiltonian grows without
bound. We note that the conservation of the Hamiltonian defined in
Eq.~\eqref{EqnHAMT} means preserving the mass-shell condition. The
unbounded growth of the error on the Hamiltonian for the RK4 method
implies that the numerical solution drift away from the mass-shell
condition, which is physically incorrect. 

\begin{figure}
\begin{centering}
\subfloat[Particle orbit]{\includegraphics[width=0.49\linewidth]{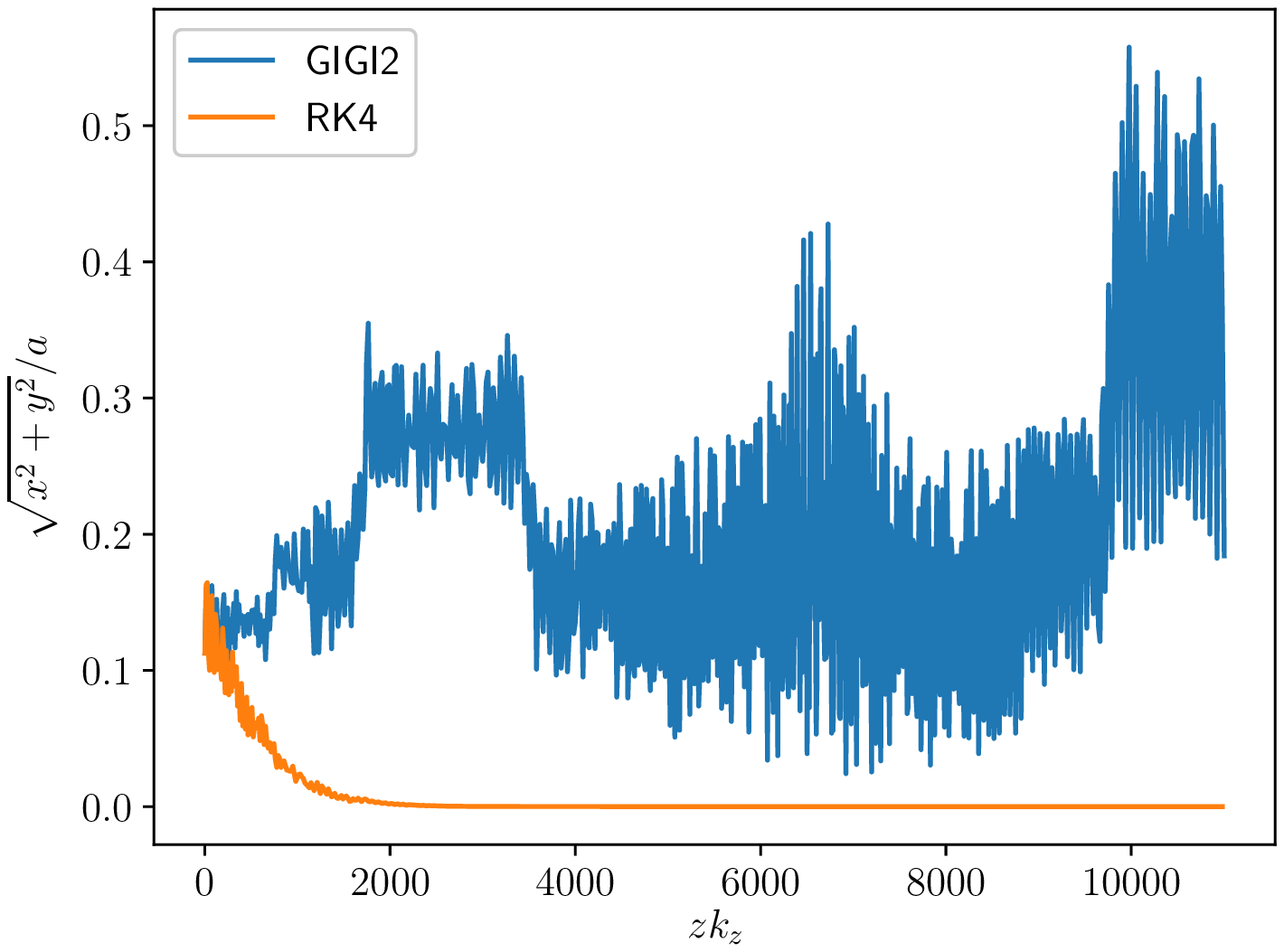}}\subfloat[Lorentz factor $\gamma$]{\includegraphics[width=0.49\linewidth]{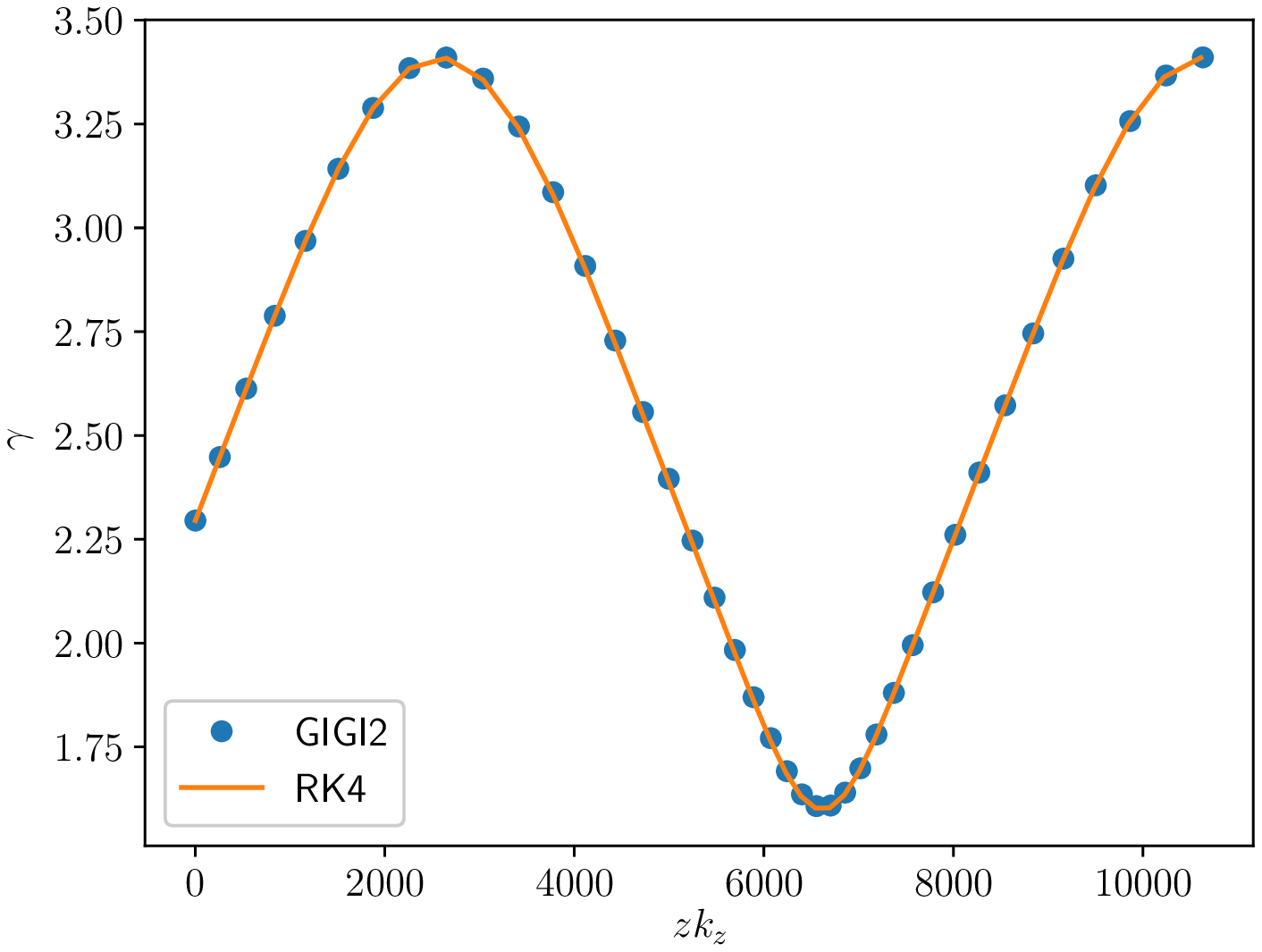}}
\par\end{centering}
\caption{Evolution of the particle orbit (a) and Lorentz factor $\gamma$ (b)
calculated by the GIGI2 and RK4 method.}
\label{FigACCENEPHI} 
\end{figure}

Next, we turn on the accelerating field. The parameters are chosen
as 
\begin{eqnarray}
\omega/\Omega & = & 0.1~,\\
k_{r}\rmc/\Omega &\approx &0.1111~,\\
\frac{\Omega\phi_{0u}q}{m_{0u}\rmc}k_{r} & = & -0.04~,
\end{eqnarray}
and the total number of time steps is 6400. Initially the phase speed
of the electric wave is the same as the speed of the particle in the
$\bfe_{z}$ direction, i.e., $\rmd z/\rmd t$. The evolution of particle
orbit and Lorentz factor $\gamma$ obtained by the GIGI2 and RK4 methods
are plotted in \FIG{FigACCENEPHI}. It shows that the particle is
accelerated at the beginning, and then decelerated and accelerated
alternatively due to the phase mis-matching and matching. The RK4
method is able to calculate correctly the energy of particle, however
it fails to compute the correct orbit in the transverse direction. 

\section{Conclusion}

\label{SecDS} In this paper, we have developed a set of explicit
high-order gauge-independent noncanonical symplectic integrators for
relativistic charged particle dynamics. These algorithms preserve
exactly a 8D noncanonical symplectic structure, and displayed long-term
accuracy and fidelity. Compared with the standard implicit symplectic
schemes for relativistic charged particles, the present schemes are
high-order and explicit. Due to their gauge-independent property,
these algorithms do not require the knowledge of vector and scalar
potentials. This is more convenient for problems where only electromagnetic
fields are given. 

\section*{Acknowledgments}
This research is supported by the National Key Research and Development
Program (2016YFA0400600, 2016YFA0400601 and 2016YFA0400602), the National
Natural Science Foundation of China (NSFC-11775219 and NSFC-11575186),
China Postdoctoral Science Foundation (2017LH002), Innovation Foundation of USTC (WK2030040096)
and the GeoAlgorithmic Plasma Simulator (GAPS) Project.

\bibliographystyle{elsarticle-num}
\bibliography{vsrel5}

\end{document}